# Fast and reliable identification of atomically thin layers of TaSe$_2$ crystals

Andres Castellanos-Gomez[1](*), Efrén Navarro-Moratalla[2], Guillermo Mokry[3], Jorge Quereda[3], Elena Pinilla-Cienfuegos[2], Nicolás Agraït[3,4], Herre S.J. van der Zant[1], Eugenio Coronado[2], Gary A. Steele[1] and Gabino Rubio-Bollinger[3](*)

[1] Kavli Institute of Nanoscience, Delft University of Technology, Lorentzweg 1, 2628 CJ Delft, The Netherlands.

[2] Instituto Ciencia Molecular (ICMol), Univ. Valencia, C/Catedrático José Beltrán 2, E-46980, Paterna, Spain.

[3] Departamento de Física de la Materia Condensada (C–III). Universidad Autónoma de Madrid, Campus de Cantoblanco, 28049 Madrid, Spain.

[4]Instituto Madrileño de Estudios Avanzados en Nanociencia IMDEA-Nanociencia. E-28049 Madrid (Spain).

a.castellanosgomez@tudelft.nl ; gabion.rubiio@uam.es

**ABSTRACT**

Deposition of clean and defect-free atomically thin two-dimensional crystalline flakes on surfaces by mechanical exfoliation of layered bulk materials has proven to be a powerful technique, but it requires a fast, reliable and non-destructive way to identify the atomically thin flakes among a crowd of thick flakes. In this work, we provide general guidelines to identify ultrathin flakes of TaSe$_2$ by means of optical microscopy and Raman spectroscopy. Additionally, we determine the optimal substrates to facilitate the optical identification of atomically thin TaSe$_2$ crystals. Experimental realization and isolation of ultrathin layers of TaSe$_2$ enables future studies on the role of the dimensionality in interesting phenomena such as superconductivity and charge density waves.

## 1. Introduction

Experimental isolation of graphene by mechanical exfoliation [1] has unleashed interest in a whole family of atomically thin materials which exhibit a variety of interesting properties ranging from topological insulator behavior to superconductivity.[2-15] While only few two-dimensional crystals have been fabricated and thoroughly characterized,[16-20] most of other possible two-dimensional (2D) crystals with attractive properties remain barely explored. For instance, layered transition metal dichalcogenides with the formula MX$_2$ (M = Mo, W, Nb, Ta or Ti and X = Se, S or Te) present a broad variety of electrical properties ranging from wide band-gap semiconductors to superconductors. As in the case of graphene, while fabrication by mechanical exfoliation from bulk crystals of these layered materials is rather simple,[2] identification of atomically thin flakes requires fast, reliable and non-destructive characterization techniques.[21-27]

The realization of ultrathin superconducting layers would enable one to employ the electric field effect to control physical properties such as the superconducting transition temperature or to study the interplay between the superconductivity and the sample dimensionality. However, among the family of transition metal dichalcogenides, the studies on atomically thin superconducting layers are scarce and mainly focused on NbSe$_2$ and TaS$_2$ crystals.[3, 4, 28] TaSe$_2$ is a good example of a layered material which has not been studied in its atomically thin form so far, while it is very interesting. In fact, in its bulk form, TaSe$_2$ is among the most studied charge density wave (CDW) systems as it shows both incommensurate and commensurate density-wave phases and a superconducting transition below 0.15 K.[29, 30]

In this work, we report the fabrication of atomically thin two-dimensional TaSe$_2$ crystals on SiO$_2$/Si wafers. We perform a combined characterization by Atomic Force Microscopy (AFM), quantitative optical microscopy and Raman spectroscopy. We also determine the optimal SiO$_2$ thickness to optically identify ultrathin TaSe$_2$ crystals. This work constitutes a necessary step towards further studies on other properties of atomically thin TaSe$_2$ sheets.

## 2. Experimental

Starting elemental materials were used as received from commercial suppliers with no further purification.

### 2.1. TaSe$_2$ single-crystal fabrication:

TaSe$_2$ crystals were synthesized from the elemental components in a two-step process. Firstly, polycrystalline TaSe$_2$ was obtained by ceramic combination of stoichiometric ratios of Ta and Se. Ta powder, 99.99% trace metals basis and Se powder, −100 mesh, 99.99% trace metals basis were used. Powdered starting materials were intimately mixed, placed inside an evacuated quartz ampoule and reacted at 900ºC during 9 days. The resulting free-flowing glittery grey microcrystals were then transformed into large single-crystals by the CVT methodology. For that purpose, 1g of TaSe$_2$ polycrystalline material together with 275 mg of I$_2$ were loaded into a 500 mm long quartz ampoule (OD: 18 mm, wall-thickness: 1.5 mm). The mixture was thoroughly stashed at one end of the ampoule and the latter was exhaustively evacuated and flame-sealed. The quartz tube was finally placed inside a three-zone split muffle where a gradient of 25 ºC was established between the leftmost load (725 ºC) and central growth (700 ºC) zones. A gradient of 25 ºC was also set between the rightmost and central regions. The temperature gradient was maintained constant during 15 days and the muffle was eventually switched off and left to cool down at ambient conditions. Millimetric TaSe$_2$ crystals were recovered from the ampoule's central zone, exhaustively rinsed with diethyl ether and stored under a N$_2$ atmosphere.

**2.2. Viscoelastic-stamp based exfoliation:**

The viscoelastic stamps employed during the TaSe$_2$ micromechanical exfoliation are based on poly (dimethil)-siloxane (PDMS) stamps, a viscoelastic material commonly used in microimprint lithography. The PDMS stamps have been cast by curing the Sylgard® 184 elastomer kit purchased from Dow Corning.[26]

**2.3. Atomic force microscopy:**

Atomic force microscopy (AFM) has been used to characterize the thickness of the fabricated flakes. A Nanotec Cervantes AFM (Nanotec Electronica) has been operated in contact mode under ambient conditions. We have selected contact mode AFM instead of dynamic modes of operation to avoid artifacts in the determination of the flake thickness.[31] The piezoelectric actuators of the AFM have been calibrated by means of a recently developed calibration method to provide a determination of the flake thickness as accurate as possible.[32]

**2.4. Optical microscopy:**

The The quantitative measurements of the optical contrast of ultrathin TaSe$_2$ flakes carried out in this work has been done with a *Nikon Eclipse LV100* optical microscope under normal incidence with a 50× objective (numerical aperture NA = 0.55) and with a digital camera *EO-1918C 1/1.8″* (from *Edmund Optics*) attached to the microscope trinocular. The illumination wavelength was selected by means of nine narrow band-pass filters (10 nm full width at half maximum FWHM) with central wavelengths 450 nm, 500 nm, 520 nm, 546 nm, 568 nm, 600 nm, 632 nm, 650 nm y 694 nm purchased from *Edmund Optics*.

**2.5. Raman spectroscopy:**

A micro-Raman spectrometer (Renishaw in via RM 2000) was used in a backscattering configuration excited with a visible laser light ($\lambda$ = 514 nm) to characterize the ultrathin TaSe$_2$ layers. The spectra were collected through a 100× objective and recorded with 1800 lines/mm grating providing the spectral resolution of ~ 1 cm$^{-1}$.

**3. Results and discussion**

Ultrathin two-dimensional crystals are fabricated out of bulk 2H-TaSe$_2$ crystals, grown by a chemical vapor growth method (see Experimental Section), using a variation of the graphene micromechanical cleavage method.[2] As previously demonstrated for other layered crystals, adhesive-free exfoliation of bulk crystals can be performed with viscoelastic silicone stamps.[12, 26, 33] The surface of the stamp containing TaSe$_2$ crystals is pressed against another clean stamp and peeled off fast to re-cleave the crystals. This maneuver is repeated several times until only faint atomically thin flakes are observed on the surface of the stamp by optical microscopy. As a final step, the crystals are transferred to an oxidized silicon wafer (300 nm thick oxide) by slightly pressing the silicone stamp against the wafer surface and subsequently peeling it off slowly. The advantage of the silicone stamp over the Scotch tape method is that one can avoid leaving adhesive traces on the wafer surface preventing the contamination of the flakes and the cantilever tip used in AFM studies of the samples.[34] In addition we have not found differences in the obtained flakes, neither for their minimal thickness nor their size, when transferred using the silicone stamp or the commonly used Scotch tape method.

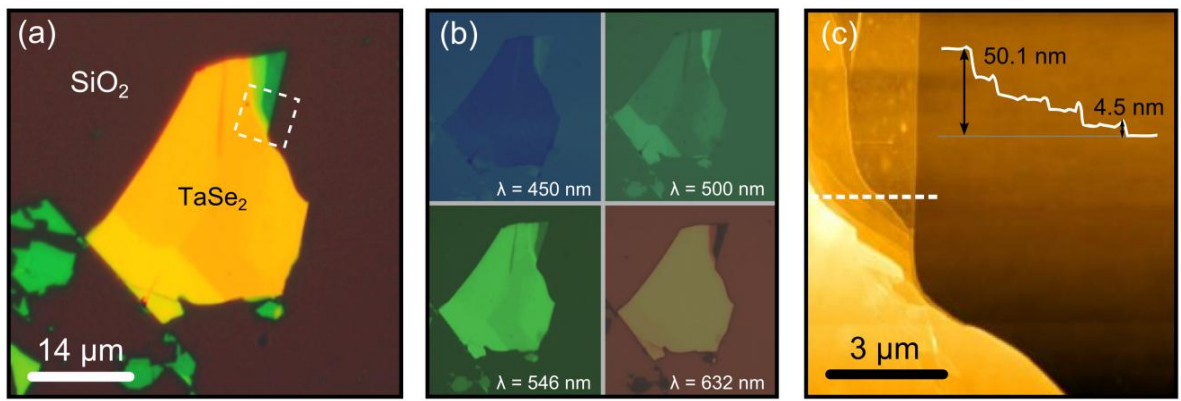

**Figure 1.** Optical and atomic force microscopy images of TaSe$_2$ nanosheets. (a) Optical micrograph under white illumination of TaSe$_2$ flakes deposited on a 300 nm SiO$_2$/Si substrate. (b) shows different optical micrographs acquired at 450 nm, 500 nm, 546 nm and 632 nm respectively by using narrow bandwidth filters. (c) AFM topography image of the region marked by a dashed rectangle in (a). A topographic line profile along the horizontal dashed line is inserted to indicate the thickness of the flakes. The root mean square roughness of the thinnest layer is 0.22 nm$^2$, while the roughness of the SiO$_2$ substrate is 0.20 nm$^2$. In some flakes the AFM topography shows a spike at the step edges, which could be attributed to rolling or folding during the deposition of the flake.

Inspection of the samples under an optical microscope enables quick identification of the flakes and their size. In addition, it is possible to get a rough estimate of the thickness of the flakes due to a light interference effect owing to the presence of the thin silicon oxide layer which makes their color thickness dependent under white light illumination (this effect is usually referred as interference color).[35] Typically, the obtained flakes with thicknesses below 40 nm are several microns wide (see Figure 1). Selected flakes are analyzed with an AFM operated in contact mode to obtain topographic images. This allows for the accurate determination of their thickness with sub-nm resolution. The number of layers present in a selected flake is obtained by simply dividing the measured thickness by the TaSe$_2$ interlayer spacing 0.64 nm [36] (see Figure 1(c) and Figure S1 of the Electronic Supplementary Material). Surprisingly, we have found that TaSe$_2$ flakes just few layers thick (5 – 15 layers) can be deposited by mechanical exfoliation and the yield rate is even higher than exfoliation of graphite (see low magnification optical images of the fabricated samples in the Figure S2 of the Electronic Supplementary Material). However, we have not yet found single or bilayer (0.7 nm or 1.4 nm thick) TaSe$_2$ flakes, at least not large enough to allow for optical identification, presumably due to a subtle balance between the interactions at the viscoelastic stamp/TaSe$_2$, interlayer TaSe$_2$ and TaSe$_2$/SiO$_2$ interfaces. Nevertheless, the expected optical contrast for a single layer TaSe$_2$ would be around -0.1 (see Figure 3 and Figure S5 of the Electronic Supplementary Material) for an optimal SiO$_2$ substrate thickness, and therefore it would be easily identifiable.

A quantitative method to determine the thickness of thin TaSe$_2$ flakes through optical microscopy would provide a fast alternative to AFM, which is a very slow technique. In this study, we have quantitatively studied the optical contrast (*C*), which depends on the flake thickness (*d*) and the illumination wavelength (*λ*):

$$C(d_1, \lambda) = \frac{I_{flake} - I_{substrate}}{I_{flake} + I_{substrate}} \qquad [1]$$

where $I_{flake}$ and $I_{substrate}$ are the reflected light intensities from the flake and the SiO$_2$ substrate, respectively. The measured contrast is compared to the contrast calculated using a model based on Fresnel's laws, which accounts for the light intensity reflected either from the substrate or from the flake [22]

$$I_{substrate}(d_1, \lambda) = \left| \frac{r_{02} + r_{23} e^{-2i\Phi_2}}{1 + r_{02} r_{23} e^{-2i\Phi_2}} \right|^2 \qquad [2]$$

$$I_{flake}(d_1, \lambda) = \left| \frac{r_{01} e^{i(\Phi_1+\Phi_2)} + r_{12} e^{-i(\Phi_1-\Phi_2)} + r_{23} e^{-i(\Phi_1+\Phi_2)} + r_{01} r_{12} r_{23} e^{i(\Phi_1-\Phi_2)}}{e^{i(\Phi_1+\Phi_2)} + r_{01} r_{12} e^{-i(\Phi_1-\Phi_2)} + r_{01} r_{23} e^{-i(\Phi_1+\Phi_2)} + r_{12} r_{23} e^{i(\Phi_1-\Phi_2)}} \right|^2$$

where the subindex 0, 1, 2 and 3 labels the different media (air, TaSe$_2$, SiO$_2$ and Si respectively), $\tilde{n}_j(\lambda) = n_j - i\kappa_j$ is the complex refractive index of the medium *j*, $d_j$ is the thickness of medium *j*, $\Phi_j = 2\pi \tilde{n}_j d_j / \lambda$ is the phase shift introduced by medium *j* and $r_{jk} = (\tilde{n}_j - \tilde{n}_k)/(\tilde{n}_j + \tilde{n}_k)$ is the amplitude of the reflected path in the interface between the media *j* and *k*.

We have analyzed the optical contrast of 30 flakes with thicknesses ranging from 3 to 36 nm (~ 4 to 56 layers), under a well-defined illumination wavelength using narrowband optical filters (see Figure 1(b)) spanning the visible spectrum (see Experimental Section). The measured optical contrast is shown in Figure 2, and accurately follows Equations [1] and [2] using the wavelength dependent complex refractive index reported in reference [37] for bulk 2H-TaSe$_2$ crystals (see Figure S3 of the Electronic Supplementary Material). The quantitative analysis of the optical contrast of the flakes under different illumination wavelengths provides a fast and reliable way to discriminate atomically thin flakes from thicker flakes. For instance, flakes thinner than 12 nm (~ 20 layers) present a negative contrast for $\lambda$ = 632 nm while thicker flakes have a positive contrast. Furthermore, flakes thinner than 6 nm (9-10 layers) can be distinguished from the rest because of their negative contrast for $\lambda$ = 600 nm.

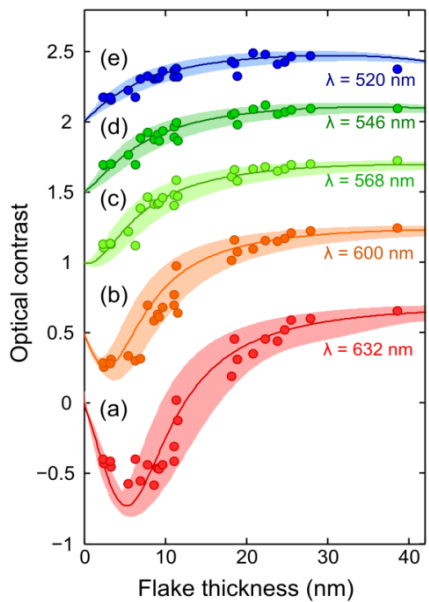

**Figure 2.** Measured optical contrast (circles) of the TaSe$_2$ flakes as a function of their thickness under the following illumination wavelengths: (a) 632 nm (red), (b) 600 nm (orange), (c) 568 nm (light green), (d) 546 nm (green) and (3) 520 nm (blue). The contrast values have been shifted 0.5, 1.0, 1.5 and 2.0 for (b), (c), (d) and (e), respectively. The solid lines correspond to the expected contrasts derived from expressions [1] and [2] using the wavelength dependent complex refractive index reported in reference [29] and considering a possible variation of *n* and *κ* of a 10 % (indicated by a shadow area around the solid line).

To determine the best SiO$_2$/Si substrate to identify ultrathin TaSe$_2$ layers, the optical contrast of a single-layer TaSe$_2$ layer has been calculated as a function of the illumination wavelength and the SiO$_2$ thickness (Figure 3(a)). We define as optimal substrates for the optical identification those with a SiO$_2$ thickness that optimize the contrast for $\lambda$ = 550 nm, which is the illumination wavelength to which the human eye attains maximum sensitivity.[38] For instance, substrates with a SiO$_2$ capping layer of 80 nm (-27 % contrast, nearly $\lambda$ independent) and 265 nm (-27 % contrast at $\lambda$ = 550 nm) optimize the optical identification of nanolayers of TaSe$_2$. Due to the widespread works on graphene, however, it is more common to find wafers with 285 nm and 90 nm SiO$_2$ thickness (which are optimized to identify graphene) in nanofabrication laboratories. Although these thicknesses are not optimal, one can identify ultrathin TaSe$_2$ flakes on those substrates. Figures 3(b) and 3(c) show the calculated optical contrast for TaSe$_2$ layers as a function of their thickness and the illumination wavelength for these two commonly used SiO$_2$/Si substrates (90 nm and 285 nm of SiO$_2$ thickness). In the Electronic Supplementary Material we also present Figures 3(b) and 3(c) recalculated for SiO$_2$ substrates that optimize the optical identification of TaSe$_2$ nanosheets (80 nm and 265 nm of SiO$_2$).

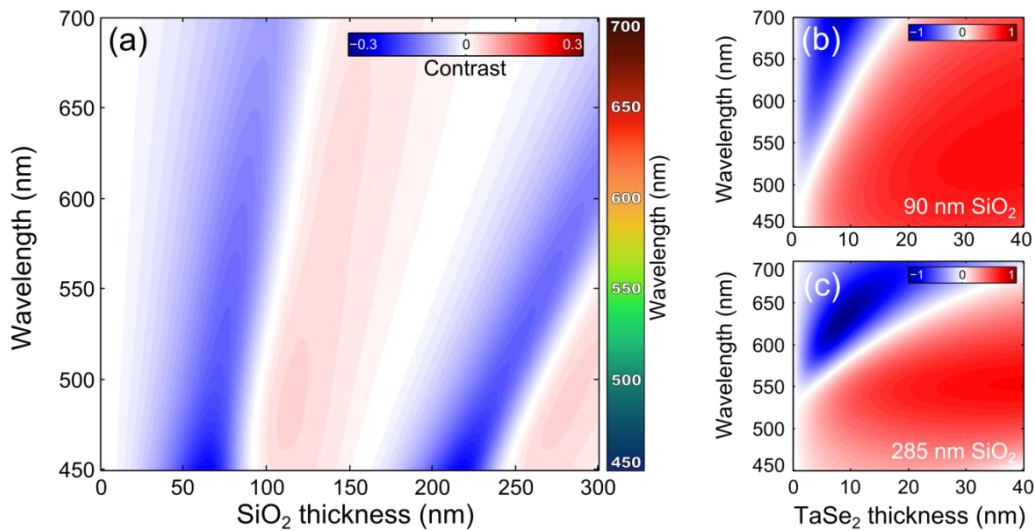

**Figure 3.** (a) Calculated optical contrast colormap for a single layer of TaSe$_2$ (0.64 nm thick) as a function of the illumination wavelength and SiO$_2$ thickness derived from eq. [1] and [2] and the complex refractive index reported for bulk 2H-TaSe$_2$ [16]. A correspondence between the apparent color and the illumination wavelength is shown in the colorbar at the right side of (a). (b) and (c) show the calculated optical contrast as a function of the wavelength and the TaSe$_2$ thicknesses for substrates with 90 nm and 285 nm SiO$_2$ thickness (which are rather standard to fabricate other 2D crystals such as graphene).

Raman spectroscopy can be considered as an alternative and complementary fast, reliable and non-destructive technique to identify ultrathin TaSe$_2$ layers. Indeed, Raman spectroscopy has been successfully employed to characterize the thickness of several atomically thin materials such as graphene [21, 39] and MoS$_2$.[40-42] Nevertheless, this technique has proven to be ineffective to detect ultrathin layers of mica [12] and, moreover, other 2D materials (such as atomically thin NbSe$_2$) are strongly damaged during the Raman spectroscopy measurements.[4] It is therefore necessary to study whether Raman microscopy can be employed to detect and to determine the thickness of ultrathin TaSe$_2$ layers.

Figure 4(a) shows the Raman Spectra measured for TaSe$_2$ flakes with thickness ranging from four layers to more than 40 layers. The spectra show five prominent peaks around 150 cm$^{-1}$, 208 cm$^{-1}$, 235 cm$^{-1}$ and 519 cm$^{-1}$ and it shows the characteristic features of 2H-TaSe$_2$.[43, 44] The Raman peak near 150 cm$^{-1}$ is due to a two-phonon scattering process.[43, 44] The peak around 208 cm$^{-1}$ (labelled as $E_{2g}^1$) corresponds to the excitation of a vibrational mode in which the Se and Ta atoms oscillate, in anti-phase, parallel to the crystal surface. The peak around 235 cm$^{-1}$ (labelled as $A_{1g}$), on the other hand, is due to the vibration of the Se atoms, in anti-phase, perpendicularly to the crystal surface while the Ta atoms are fixed.[43, 44] Finally, the peak at 519 cm$^{-1}$ is due to the vibration of the lattice of the silicon substrate underneath.

The position of the $E_{2g}^1$ and $A_{1g}$ peaks as a function of the TaSe$_2$ layer thickness is presented in Figure 4(b). Within a classical model for coupled harmonic oscillators,[45] both the $E_{2g}^1$ and $A_{1g}$ modes are expected to shift to lower frequencies as the number of layers decreases because the interlayer van der Waals interaction decreases and so the effective restoring forces acting on the atoms. However, we observed that the $E_{2g}^1$ peak shifts to higher frequencies as the number of layers decreases and the position of the $A_{1g}$ peak shows no clear shift within the experimental resolution, in agreement with previous experiments in intercalated TaSe$_2$ samples.[44] The observed shift of the $E_{2g}^1$ peak has been also observed in another transition metal dichalcogenide, the MoS$_2$, and it has been explained by an enhancement of the dielectric screening of the long-range Coulomb interaction between the effective charges with increasing number of layers which reduces the overall restoring force on the atoms.[46]

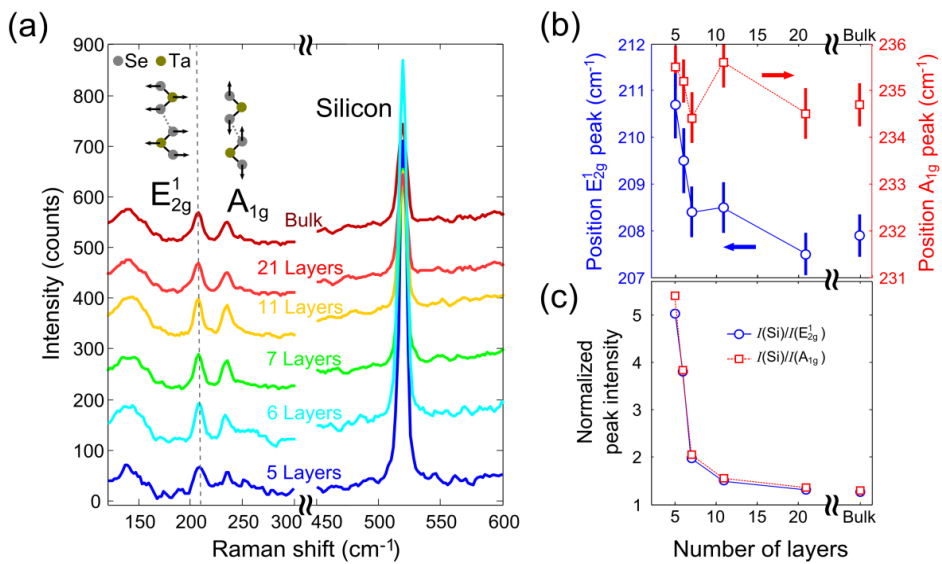

**Figure 4:** (a) Raman spectra of TaSe$_2$ flakes with thicknesses ranging from five layers to > 40 layers (bulk). (b) Position of the $E_{2g}^1$ and $A_{1g}$ Raman modes. (c) Ratio between the maximum intensities of the silicon and the $E_{2g}^1$ and $A_{1g}$ peaks.

The monotonic thickness dependence of the $E_{2g}^1$ peak position can be employed to determine the number of layers of TaSe$_2$ flakes, as it has been previously employed for other materials such as MoS$_2$.[40, 42] The signal-to-noise ratio of this frequency shift, however, would make this determination rather inaccurate for flakes thinner than 7 layers (1-2 layer of uncertainty) and even impractical for thicker flakes. Nevertheless, as for other dichalcogenides (such as MoS$_2$)[6] the E$_{2g}^1$ shift may be even larger for single and bilayer TaSe$_2$ helping to identify them. An alternative procedure to determine the number of layers of TaSe$_2$ flakes relies on measuring the ratio between the intensities of the silicon peak and the E$_{2g}^1$ and A$_{1g}$ peaks.[47] Figure 4(c) shows that $I$(Si)/$I$(E$_{2g}^1$) and $I$(Si)/$I$(A$_{1g}$) depend monotonically and strongly with the number of layers. The thickness of flakes thinner than 20 layers can be determined with less than 1 layer uncertainty. Notice that for substrates with a different SiO$_2$ thickness, the thickness dependence of the ratios $I$(Si)/$I$(E$_{2g}^1$) and $I$(Si)/$I$(A$_{1g}$) will be different but after a calibration measurement (as Figure 4(c)) they can be used to accurately determine the number of layers of TaSe$_2$ flakes.

To avoid laser-induced modification of the samples,[48] all spectra were recorded with a power level P = 0.5 – 1 mW and an accumulation time of one second. Interestingly, we have found that the laser-induced damage of TaSe$_2$ flakes can be monitored by measuring the Raman spectrum. Figure 5 shows the Raman spectra measured in a five layers TaSe$_2$ flake using one sec. (blue line) and two sec. (red lines) accumulation time. Both the Raman spectrum and the topography of the flake demonstrate that the flake is modified, possibly due to laser-induced oxidization, when two sec. of accumulation time is employed. A peak around 260 cm$^{-1}$ appears together to the laser-induced modification of the topography of the TaSe$_2$ flake. Therefore, the appearance of this peak can be used to regulate the laser intensity in order to avoid laser-induced oxidation of the flakes.

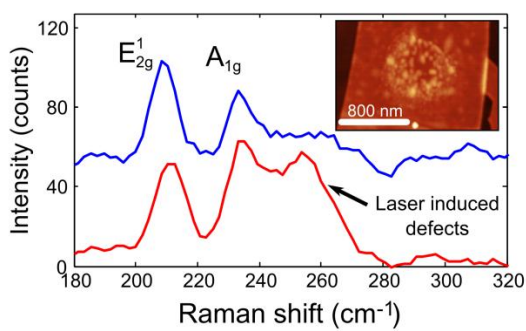

**Figure 5:** Raman spectra of a TaSe$_2$ flake with 5 layers measured with 1 sec. (blue line) and 2 sec. accumulation time (red line). (Inset) Atomic Force Microscopy topography image of the flake after the 2 sec. acquisition time Raman spectrum, shows that the layer has been modified by the laser beam.

**4. Conclusions**

In summary, atomically thin TaSe$_2$ sheets have been fabricated by mechanical exfoliation of bulk 2H-TaSe$_2$ single crystals grown by chemical vapor transport method. We demonstrated that optical microscopy and Raman spectroscopy can be used to identify atomically thin TaSe$_2$ crystals and to distinguish them from thicker crystals. The optimal SiO$_2$ thickness to facilitate the optical identification of these atomically thin flakes on SiO$_2$/Si wafers has been also calculated. Additionally, we have demonstrated that Raman spectroscopy can be used to identify ultrathin TaSe$_2$ flakes and to monitor any possible laser-induced damage.This work can be thus considered as a necessary first step towards the deeper study of ultrathin TaSe$_2$.

**Acknowledgements**

This work was supported by the Spanish MICINN/MINECO (projects MAT2011-25046, MAT2011-22785 and CONSOLIDER-INGENIO-2010 on 'Nanociencia Molecular', CSD-2007-00010), the Comunidad de Madrid (program Nanobiomagnet S2009/MAT-1726), the Generalidad Valenciana (PROMETEO and ISIC programs) and the European Union (Projects RODIN and ELFOS).

**Electronic Supplementary Material**

# Fast and reliable identification of atomically thin layers of TaSe$_2$ crystals

Andres Castellanos-Gomez[1](*), Efrén Navarro-Moratalla[2], Guillermo Mokry[3], Jorge Quereda[3], Elena Pinilla-Cienfuegos[2], Nicolás Agraït[3,4], Herre S.J. van der Zant[1], Eugenio Coronado[2], Gary A. Steele[1] and Gabino Rubio-Bollinger[3](*)

[1] Kavli Institute of Nanoscience, Delft University of Technology, Lorentzweg 1, 2628 CJ Delft, The Netherlands.
[2] Instituto Ciencia Molecular (ICMol), Univ. Valencia, C/Catedrático José Beltrán 2, E-46980, Paterna, Spain.
[3] Departamento de Física de la Materia Condensada (C–III). Universidad Autónoma de Madrid, Campus de Cantoblanco, 28049 Madrid, Spain.
[4] Instituto Madrileño de Estudios Avanzados en Nanociencia IMDEA-Nanociencia. E-28049 Madrid (Spain).

**Characterization of atomically thin TaSe$_2$ crystals**

Figure S1 shows the same information as Figure 1 in the main text for other two TaSe$_2$ flakes. Both the optical and atomic force microscopy characterization of the flakes is shown.

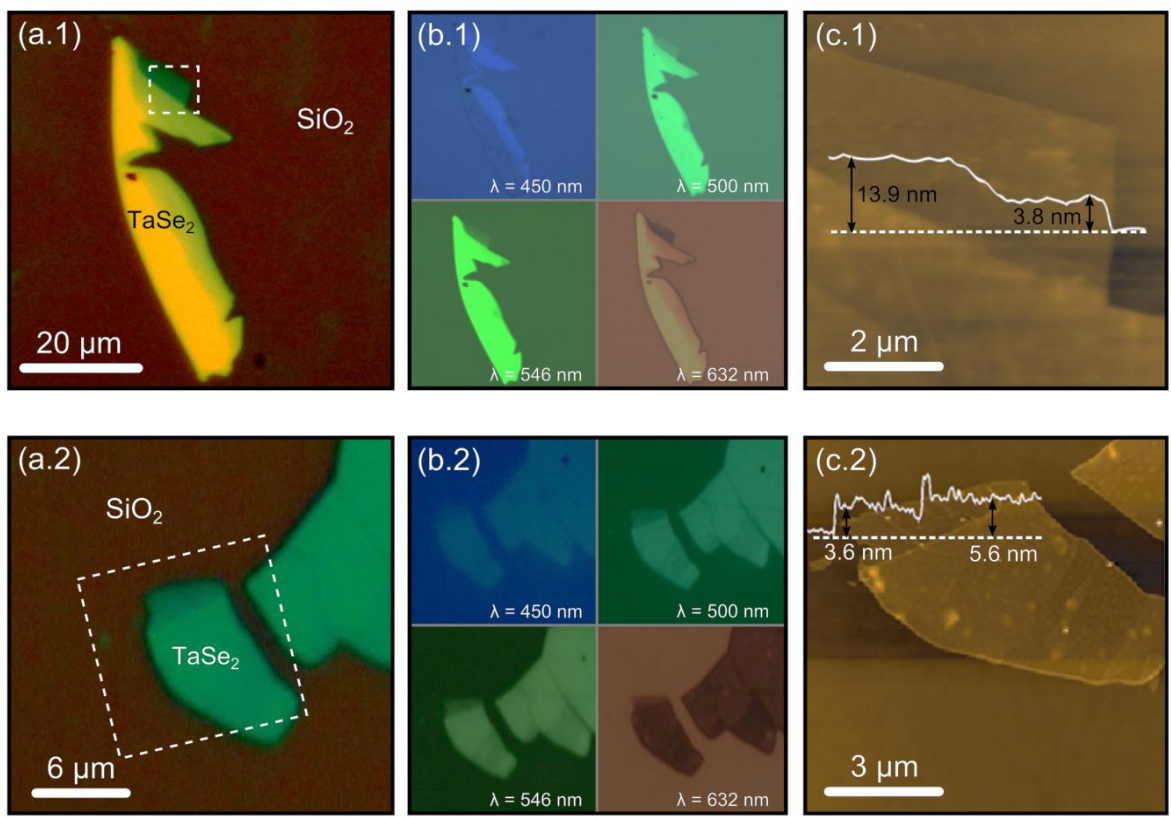

Figure S1. Optical and atomic force microscopy images of different $TaSe_2$ nanosheets. (a) Optical micrographs under white illumination of $TaSe_2$ flakes deposited on a 300 nm $SiO_2$/Si substrate. (b) shows different optical micrographs acquired at 450 nm, 500 nm, 546 nm and 632 nm respectively by using narrow bandwidth filters. (c) AFM topography image of the region marked by a dashed rectangle in (a). A topographic line profile along the horizontal line is inserted to indicate the thickness of the flakes.

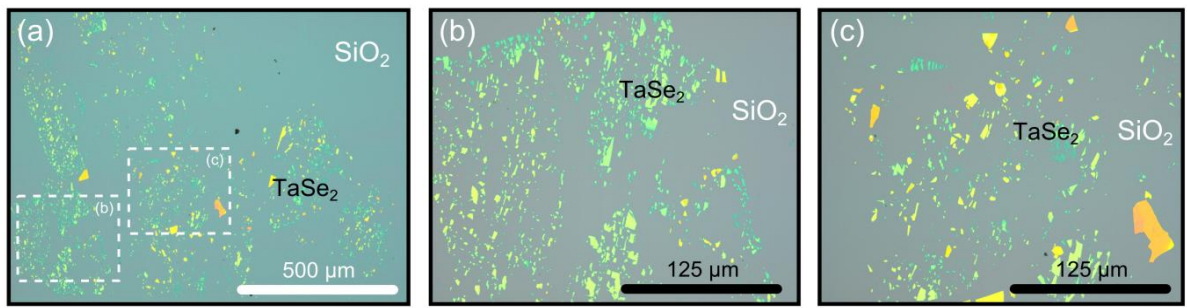

Figure S2. Low magnification optical images of few-layer $TaSe_2$ crystals exfoliated onto a 300 nm $SiO_2$/Si substrate. Greenish flakes are typically thinner than 10-15 layers. (b) and (c) show a close up image in the areas marked by the dashed rectangle in (a).

**Refractive index of $TaSe_2$**

For the calculation of the optical contrast of atomically thin $TaSe_2$ crystals (see Figure 2 and Figure 3 of the main manuscript) we have employed the complex refractive index reported for bulk 2H-$TaSe_2$[1]. In Ref. [1], however, the two components of the complex dielectric permittivity are shown. In Figure S3 we show the complex refractive index calculated from the dielectric permittivity values reported in Ref.[1].

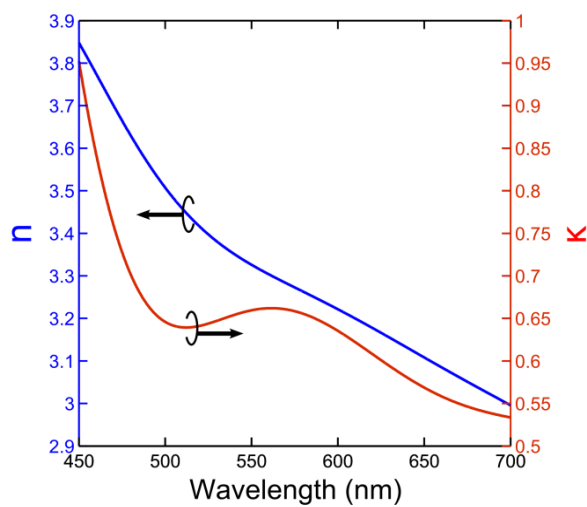

Figure S3. Complex refractive index (real part n and absorption κ) of 2H-$TaSe_2$ obtained from the dielectric permittivity values reported in Ref. [1].

**Optimal dielectric thickness to facilitate the optical identification**

The optical contrast yielded by a single layer $TaSe_2$ sheet has been calculated using Expressions [1] and [2] from the main text and the refractive index reported for bulk 2H-$TaSe_2$ [1] (see Figure S4). We have considered the case of different dielectric layers such as: $SiO_2$ (a), PMMA (b), $Al_2O_3$ (c), $HfO_2$ (d) and $Si_3N_4$ (e). The optimal dielectric thickness to identify ultrathin layers of TaSe$_2$ has been marked in Figure S4 with white circles.

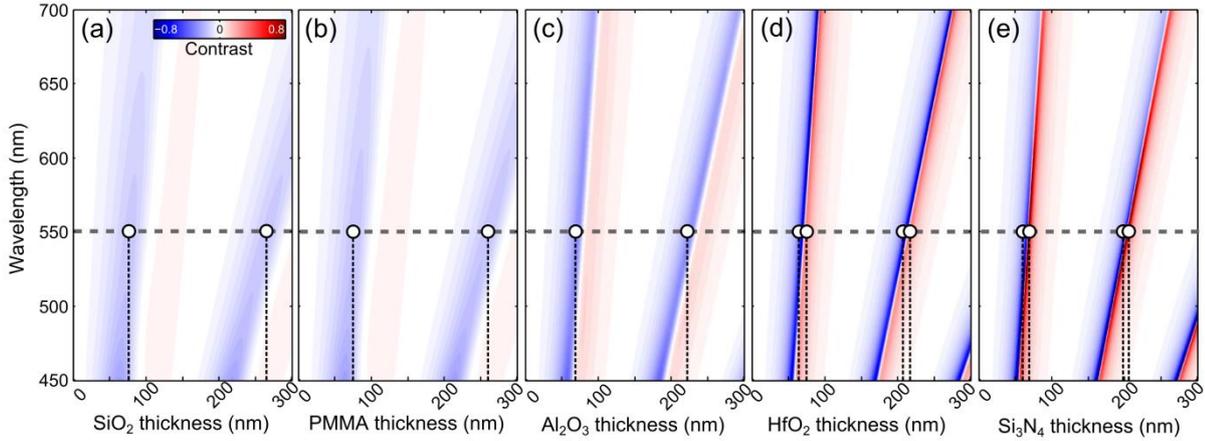

Figure S4. Optical contrast for a single layer TaSe$_2$, 0.64 nm thick, (colormap) as a function of the wavelength and dielectric thickness calculated with eq. (1) and (2) and the complex refractive index reported for bulk 2H-TaSe$_2$ [1].

For SiO$_2$, for instance, the optimal thicknesses to facilitate the optical identification are 80 nm and 265 nm. In Figure S5 we present the calculated optical contrast as a function of the illumination wavelength and the TaSe$_2$ thickness for these SiO$_2$/Si substrates.

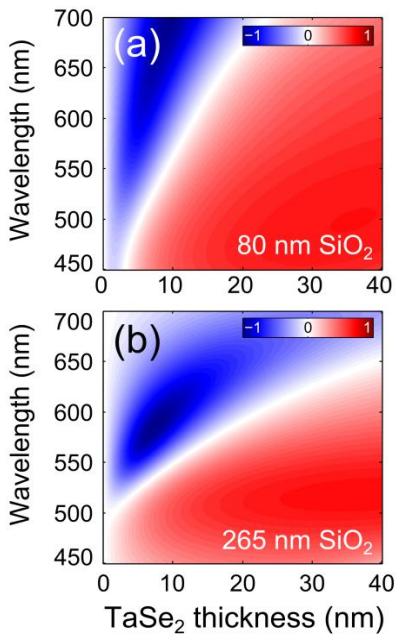

Figure S5. (a) and (b) show the calculated optical contrast as a function of the wavelength and the TaSe$_2$ thicknesses for substrates with 80 nm and 265 nm SiO$_2$ thickness (which are optimal to identify ultrathin TaSe$_2$ on SiO$_2$/Si wafers).

**Characterization of the laser-induced defects**

As pointed out in Figure 5 of the main text, prolonged or high intensity Raman spectroscopy measurements on TaSe$_2$ flakes causes laser-induced defects. These defects can be easily identified because they change drastically the optical contrast of the TaSe$_2$ flakes. Figure S6a shows an optical image of TaSe$_2$ several flakes. The central flake has been subjected to a prolonged Raman spectroscopy image which damaged the flake as can be seen form the change in contrast inside the black frame in Figure S6a. Figure S6b shows the AFM topography of the damaged region.

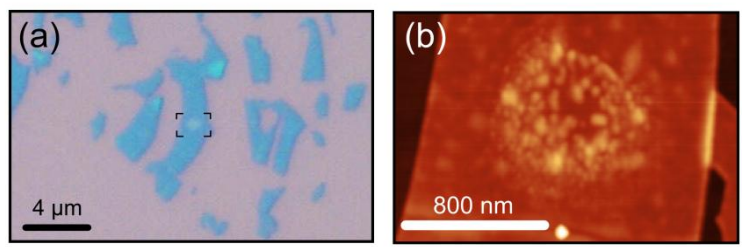

Figure S6. (a) Optical image of a TaSe$_2$ after an intrusive Raman spectroscopy measurement. The laser-induced defects are clearly visible as a drastic change in optical contrast (see region within the black frame). (b) AFM topography image of the region within the black frame in (a) showing that the region modified during the Raman spectroscopy measurement.

As seen in Figure S6, although the optical contrast drastically changed, the thickness of the flake is scarcely modified. This indicates that the laser-induced damage of the flake alters its optical contrast. This is in agreement with laser-induced oxidation as tantalum oxide is expected to have a lower optical absorption coefficient than TaSe$_2$, explaining the reduced optical contrast. Therefore, optical contrast can be used to determine whether a Raman spectroscopy measurement has damaged the sample or not.

Figure S7 shows another example where a high power laser is scanned over part of the surface of a TaSe$_2$ flake to intentionally damage it. The optical images before (a) and after (b) clearly show that one can identify the laser-modified part just by the difference in optical contrast. The AFM topography in Figure S7c shows a slight decrease of the thickness of the region modified with the laser. Note that this small reduction in the thickness is not sufficient to explain the drastic reduction of the optical contrast in the scanned area.

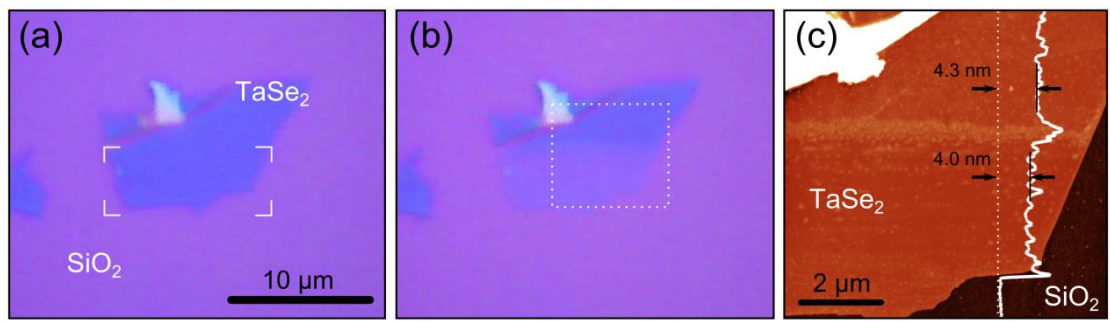

Figure S7. (a) and (b) are optical images of a TaSe$_2$ before and after scanning a high-power laser (10 mW, 500 nm spot size, scanning steps of 500 nm with a waiting time of 0.1 sec between steps). The laser scanning has been carried out in the region denoted by the white frame in (a). (c) shows the AFM topography of the region marked by the dotted square in (b). A topographic line profile has been included in (c).

In order to obtain more information about the nature of the laser-induced damage on the TaSe$_2$ flakes we have carried out Raman spectroscopy measurements on a TaSe$_2$ flake while increasing the laser power (Figure S8). The flake was deposited onto a gold electrode to avoid the strong signal from the silicon underneath (with intense peaks around 950 cm$^{-1}$, 520 cm$^{-1}$ and 300 cm$^{-1}$), which hampers the identification of some of the laser-induced features in the Raman spectra. For low power laser, the Raman spectra show the E$_{2g}^1$ and A$_{1g}$ peaks (see Figure 4 of the main text) and a background signal. For increasing power, however, a feature around 255 cm$^{-1}$ starts to develop (as seen in Figure 5 of the main text) and becomes more and more pronounced as the exposure to the laser increases. This Raman peak matches the energy of the most intense Raman peak for crystalline Ta$_2$O$_5$ in that part of the spectrum [2]. Furthermore, together with the appearance of the peak at 255 cm$^{-1}$, two broad peaks around 1350 cm$^{-1}$ and 1550 cm$^{-1}$ develop. These peaks correspond to a photoluminescence emission around 550 nm to 560 nm. Note that photoluminescence emission of similar characteristics has been observed for Ta$_2$O$_5$ of different forms (nanorods, nanoblocks, etc) [3,4].

In summary, the observed change in the optical contrast, presence of a Raman peak at 255 cm$^{-1}$ and photoluminescence emission around 550-560 nm suggest that the laser-induced defect is probably due to the oxidation of tantalum in a form of Ta$_2$O$_5$. Nevertheless, more work would be necessary to study this oxidation process in more detail, which is out of the scope of the current work.

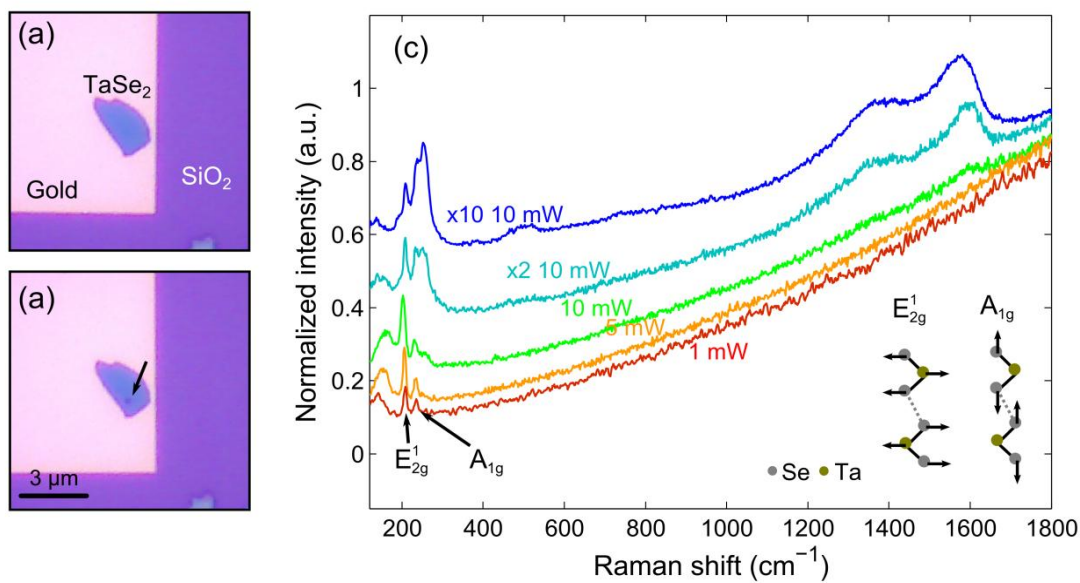

Figure S8. (a) and (b) are optical images of a TaSe$_2$ deposited on a gold electrode before and after performing several Raman spectroscopy measurements with different laser power. (c) Raman spectra measured on the TaSe$_2$ with 1mW (1 sec. accumulation), 5mW (1 sec. accumulation), 10 mW (1 sec. accumulation), 10 mW (2 sec. accumulation) and 10 mW (10 sec. accumulation) laser power.